\documentclass[conference]{IEEEtran}

\usepackage{paco-notations}
\usepackage{supertech-llncs-light}
\usepackage{cleveref}

\usepackage{amsmath,amssymb,amsfonts}
\usepackage{algorithmic}
\usepackage{graphicx}
\usepackage{textcomp}
\usepackage{xcolor}
\usepackage[numbers]{natbib}
\def\BibTeX{{\rm B\kern-.05em{\sc i\kern-.025em b}\kern-.08em
    T\kern-.1667em\lower.7ex\hbox{E}\kern-.125emX}}
    
\begin{document}
\title{A Reexamination of the COnfLUX 2.5D LU Factorization Algorithm}

\author{\IEEEauthorblockN{Yuan Tang}
\IEEEauthorblockA{\textit{School of Computer Science, School of Software} \\
\textit{Fudan University}\\
Shanghai, P.R.China \\
yuantang@fudan.edu.cn}
}
\punt{
\author{\IEEEauthorblockN{Anonymous Author(s)}
}
}
\maketitle

\begin{abstract}

This article conducts a reexamination of the research conducted by Kwasniewski et al., focusing on their adaptation of the 2.5D LU factorization algorithm with tournament pivoting, known as \func{COnfLUX}. Our reexamination reveals potential concerns regarding the upper bound, empirical investigation methods, and lower bound, despite the original study providing a theoretical foundation and an instantiation of the proposed algorithm. This paper offers a reexamination of these matters, highlighting probable shortcomings in the original investigation. Our observations are intended to enhance the development and comprehension of parallel matrix factorization algorithms.
\end{abstract}

\punt{

\ccsdesc[500]{Theory of computation~Shared memory algorithms}
\ccsdesc[500]{Theory of computation~Divide and conquer}
}
\begin{IEEEkeywords}
COnfLUX algorithm,
communication bandwidth,
LU factorization,
2.5D algorithm
\end{IEEEkeywords}

\secput{intro}{Introduction}
Matrix factorizations play a vital role in various scientific computations. In the realm of high-performance computing, there has been significant interest in developing optimized algorithms for factorizations. Kwasniewski et al. \cite{KwasniewskiKaBe21} have made a remarkable contribution to this field with their work on the \func{COnfLUX} algorithm. This variant of the 2.5D LU factorization algorithm with tournament pivoting is presented in their paper, which includes a theoretical framework, experimental validation, and the derivation of a matching lower bound. 

However, after conducting a careful reexamination, we have identified several potential issues in their paper. The focus of this article is to conduct a technical re-examination of the \func{COnfLUX} algorithm and its associated analyses, aiming to address these identified concerns. Our analysis specifically focuses on scrutinizing the estimation of the upper bound in the \func{COnfLUX} algorithm, examining its experimental methods, and the corresponding lower bound. 

\paragrf{Identified Issues}

Through a careful review of the \func{COnfLUX} algorithm and its accompanying analyses in the original paper, we have pinpointed potential issues in the following areas, listed in order of significance:

\begin{enumerate}
    \item The upper bound: We have observed a discrepancy between the authors' analyses and the actual costs incurred by the algorithm. Specifically, the utilization of a 1D decomposition for certain regions (for panel factorization and \proc{TRSM}) in the algorithm may not fully harness the communication capabilities of \emph{all} processors, resulting in an underestimation of the communication bandwidth cost.
    \item The empirical methods: Upon examining the original code base \footnote{Snapshot taken on May 29, 2023 from \href{https://github.com/eth-cscs/conflux}{https://github.com/eth-cscs/conflux}}, we have discovered that the authors only tested certain processor grid configurations and did not evaluate the \co{} configurations stated in the paper. This discrepancy has the potential to affect the validity of the claim regarding the proposed \func{COnfLUX} algorithm's communication optimality.
    \item The lower bound: The lower bound derivation may oversimplify the matter by not considering the fact that in parallel computation, the total amount of I/O operations typically increases proportionally to the number of processors, which is usually asymptotically larger than in sequential case.
\end{enumerate}

    \begin{figure}[htbp]
		\centering
			\fbox{\includegraphics[width=\linewidth]{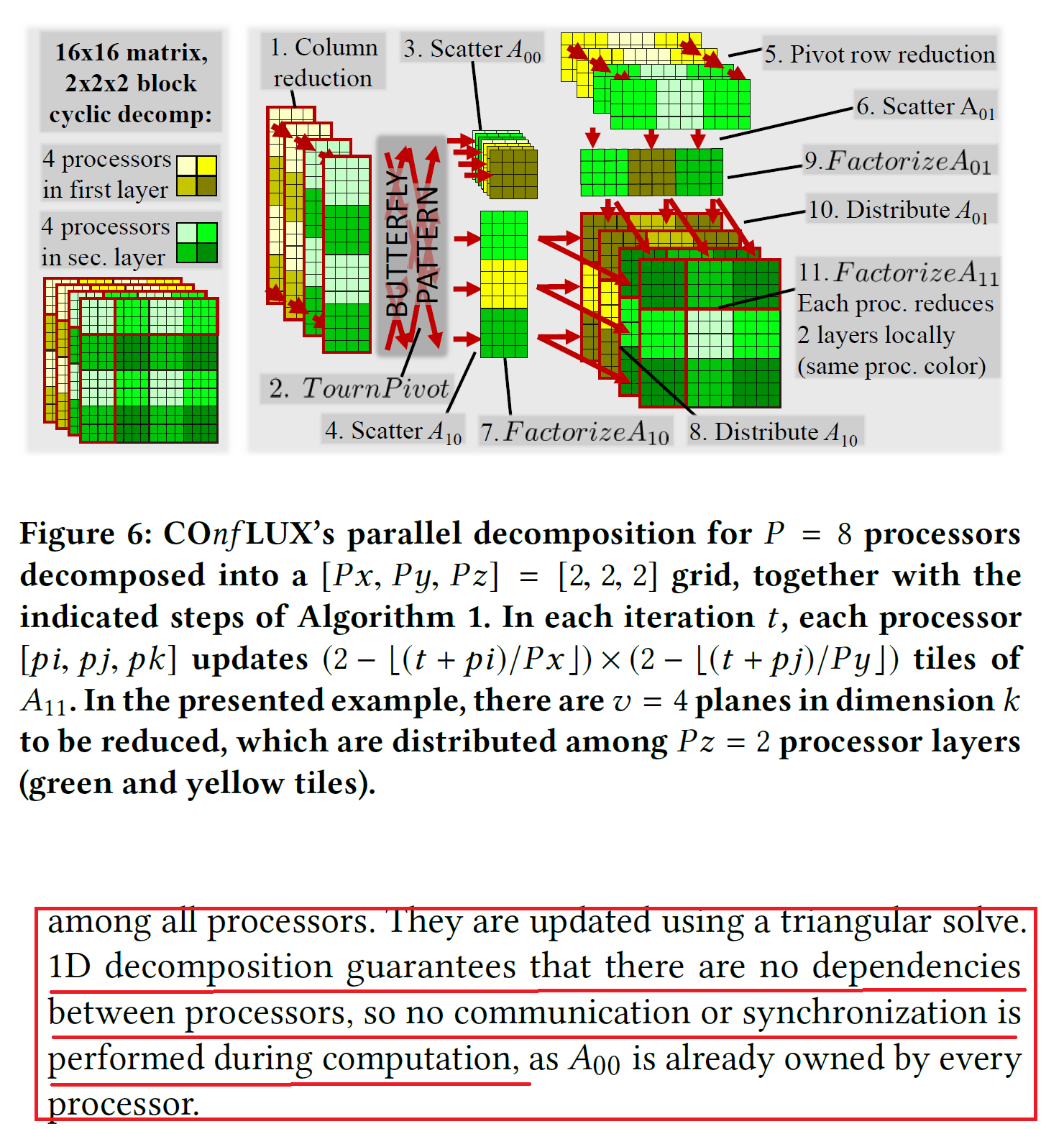}}
			 \caption{Description of using 1D decomposition for the $A_{10}$ and $A_{01}$ regions of LU -- in Sect. 7.2 of original paper \cite{KwasniewskiKaBe21}}
			 \label{fig:conflux-1d-decomp}
	\end{figure}
 
	\begin{figure}[htbp]
		\centering
			\fbox{\includegraphics[width=\linewidth]{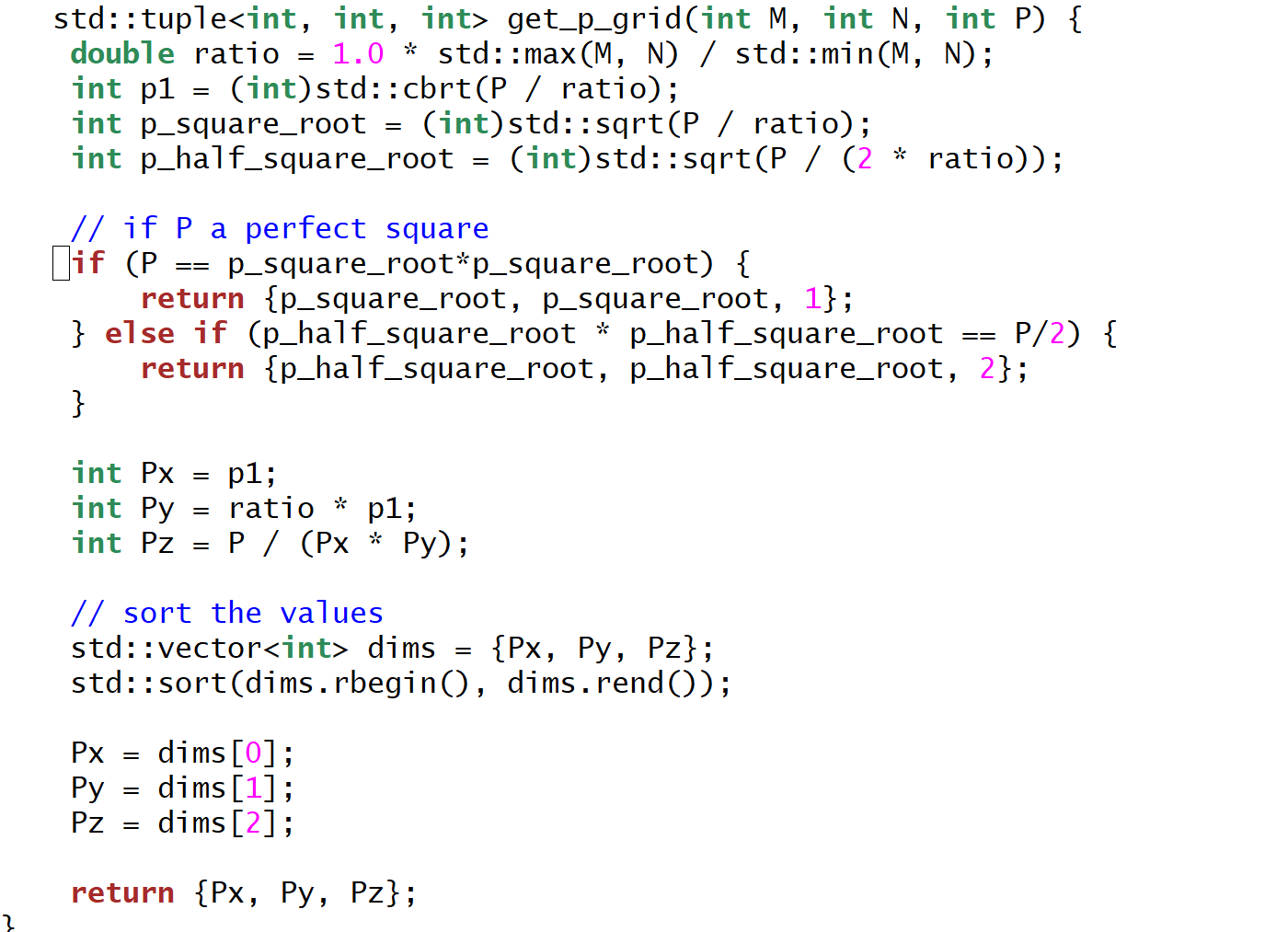}}
			\caption{Snapshot from original ``lu\_params.hpp'' of code base showing that its processor grid setting is $\sqrt{p} \times \sqrt{p} \times 1$ or $\sqrt{p/2} \times \sqrt{p/2} \times 2$}
		\label{fig:conflux-processor-grid}
	\end{figure}
 
	\begin{figure}[htbp]
		\centering
			\fbox{\includegraphics[width=\linewidth]{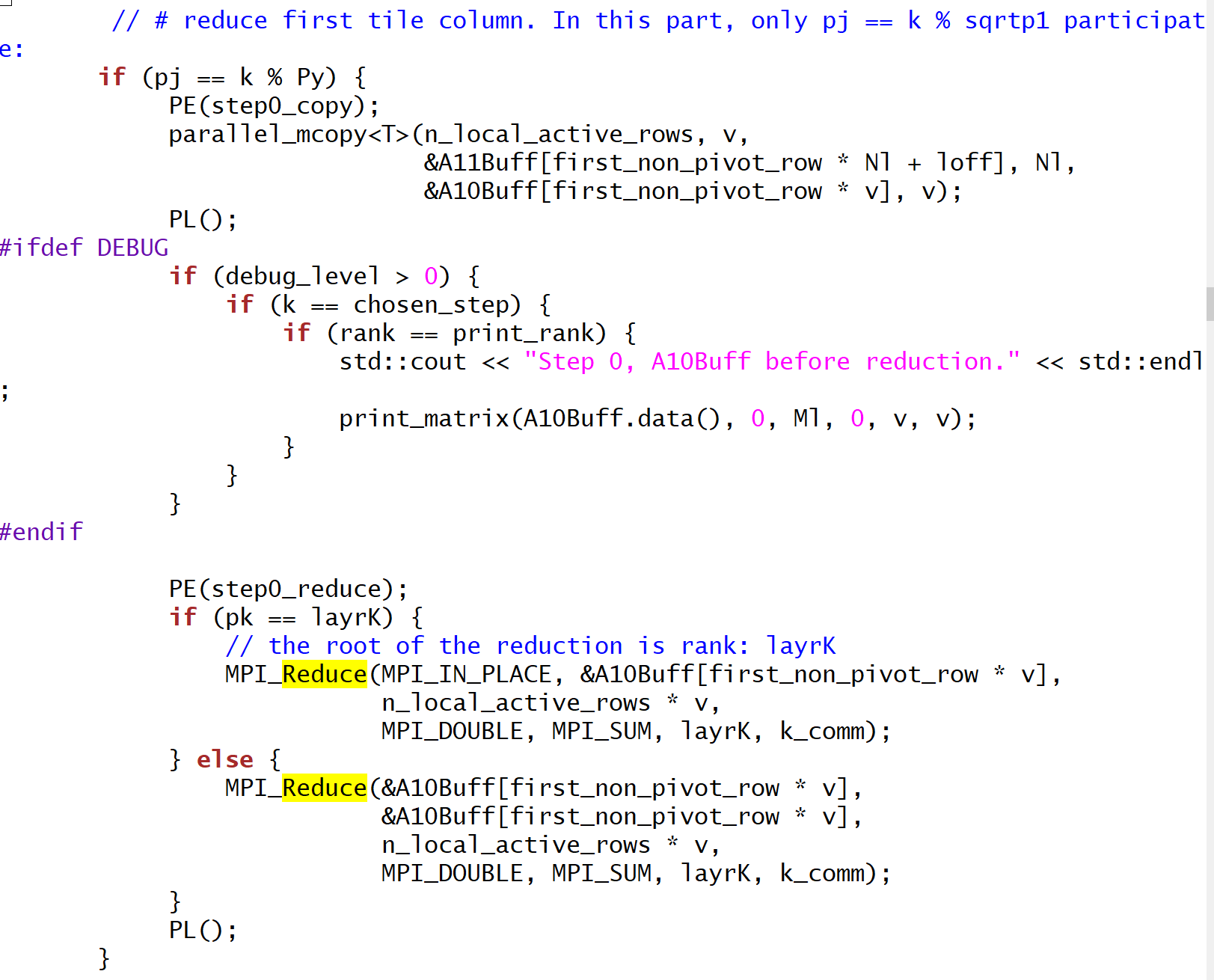}}
		\caption{Snapshot from original ``conflux\_opt.hpp'' of code base showing that \func{COnfLUX} employs at most $\id{pi} \cdot \id{pk} = p^{1/2}_1 c = O(\sqrt{p})$ processors in the reduction operations of the $A_{10}$ and $A_{01}$ regions.}
		\label{fig:conflux-panel-reduction}
	\end{figure}
 
    \begin{figure}
        \centering
        \fbox{\includegraphics[width=\linewidth]{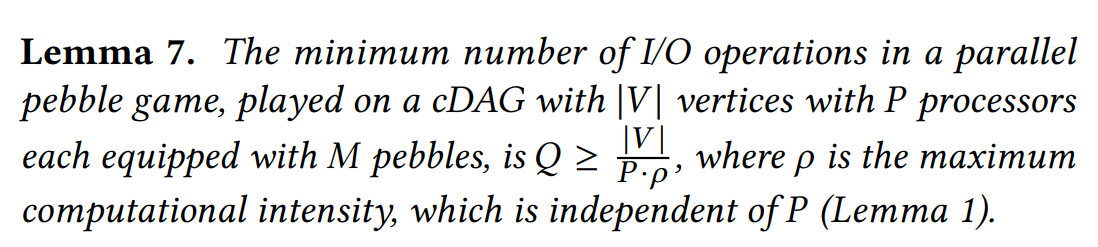}}
        \caption{Lemma 7 in Sect. 5 of original paper}
        \label{fig:conflux-lemma-7}
    \end{figure}

    \begin{figure}
        \centering
        \fbox{\includegraphics[width=\linewidth]{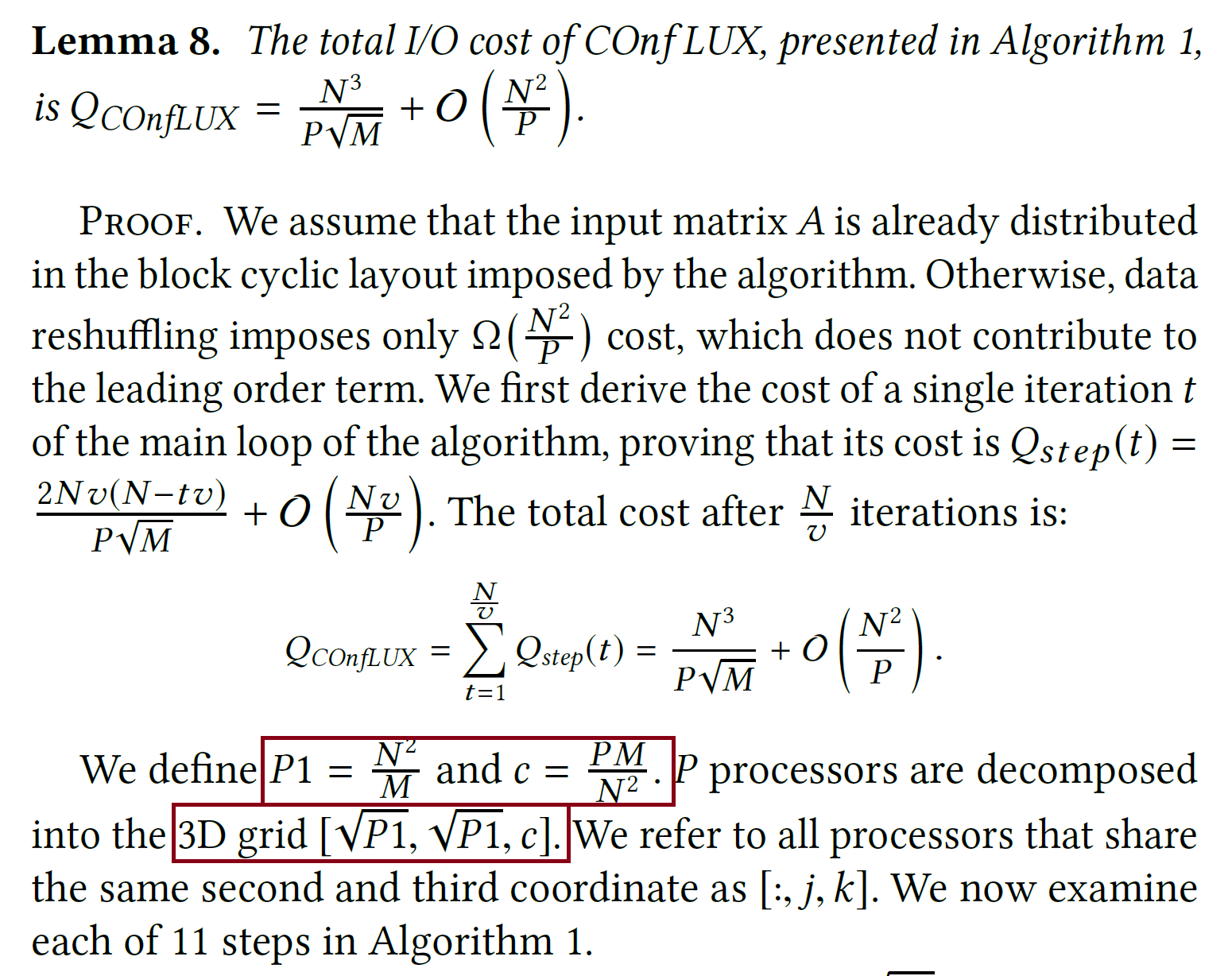}}
        \caption{Lemma 8 in Sect. 7.4 of original paper reveals the processor grid configuration is $p_1^{1/2} \times p_1^{1/2} \times c$.}
        \label{fig:conflux-lemma-8}
    \end{figure}

    \begin{figure}
        \centering
        \fbox{\includegraphics[width=\linewidth]{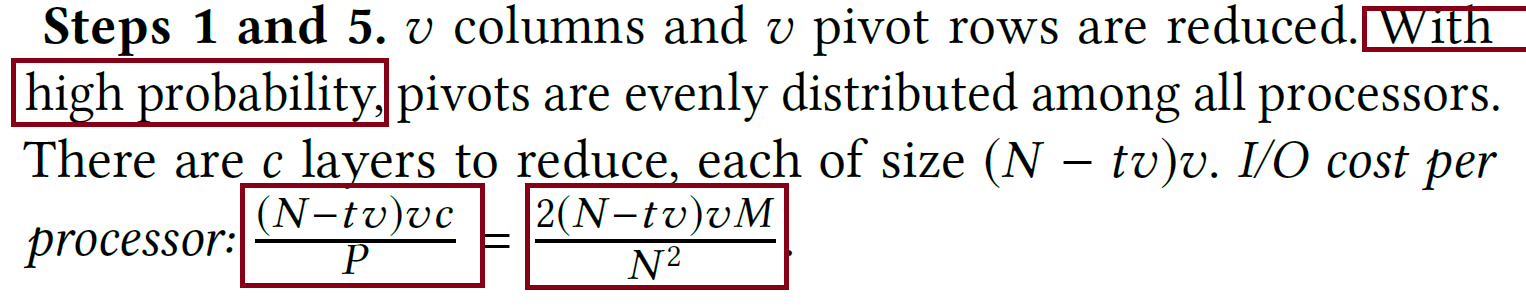}}
        \caption{The I/O cost calculation of steps 1 and 5 in original Lemma 8 shows that it distributes the I/O cost over all $p$ processors, rather than the $p_1^{1/2}c$ processors actively involved.}
        \label{fig:conflux-io-steps-1-5}
    \end{figure}
    
    \begin{figure*}
        \centering
        \fbox{\includegraphics[width=\linewidth]{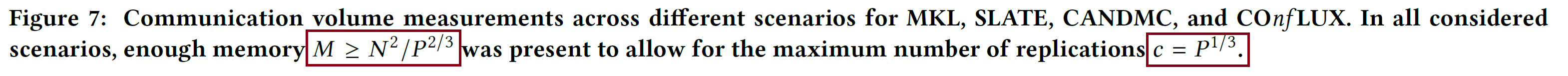}}
        \caption{The caption of Figure 7 of original paper \cite{KwasniewskiKaBe21} implies that the \co{} setting is $p_1^{1/2} \times p_1^{1/2} \times c = p^{1/3} \times p^{1/3} \times p^{1/3}$}
        \label{fig:conflux-cap-fig-7}
    \end{figure*}
    
\secput{issues-ub-conflux}{Issues in the Communication Bandwidth Upper Bound}

Kwasniewski et al. \cite{KwasniewskiKaBe21} introduced \func{COnfLUX}, a variant of the 2.5D LU factorization algorithm with tournament pivoting. A distinction from the algorithm by Solomonik and Demmel \cite{SolomonikDe11} is the use of a 1D decomposition for panel factorization in the $A_{10}$ region and \proc{TRSM} in the $A_{01}$ regions, as shown in \afigref{conflux-1d-decomp}. The aim of using a 1D algorithm is to remove dependencies between processors, eliminating inter-processor communication during these computations \footnote{The \func{COnfLUX} algorithm uses tournament pivoting, which differs from the partial pivoting in LAPACK's \func{GETF2} routine}.

However, this approach introduces ambiguity regarding how \emph{all} $p$ processors could possibly contribute to the reduction operation essential for panel factorization and \proc{TRSM}, a critical step for achieving bandwidth (communication volume) optimality along a critical path in the 2.5D LU algorithm by Solomonik and Demmel \cite{SolomonikDe11}. By default, this paper focuses on the complexity of ``communication bandwidth'' and its bound along the critical path, unless stated otherwise. In this context, we refer to the Kwasniewski et al.'s paper and code base on the \func{COnfLUX} algorithm as the "original paper / code" \cite{KwasniewskiKaBe21}.

Through analysis of the original paper and code base \footnote{Snapshot of the code base was taken on May 29, 2023, available at \href{https://github.com/eth-cscs/conflux}{https://github.com/eth-cscs/conflux}}, the bandwidth cost for the reduction in $A_{10}$ and $A_{01}$ regions is re-examined as follows. 

The original paper calculates this cost in ``steps 1 and 5'' of Section 7.4, as shown in \figref{conflux-io-steps-1-5}, and is formulated as follows:

\begin{align}
\frac{(n - tv)vc}{p} = \frac{2 (n - tv) v \mM}{n^2} \label{eq:conflux-reduction-cost}
\end{align}

Here, $v$ is the panel width, $t$ the iteration count, and $c$ the layer count, with a processor grid of $p = p^{1/2}_1 \times p^{1/2}_1 \times c$ (refer to \figref{conflux-lemma-8}). So equation \eqref{conflux-reduction-cost} can be explained as follows: $v$ columns and $v$ pivot rows are reduced, assuming that the pivots are evenly distributed among the processors over $c$ layers, each of size $(n - tv)v$. However, the issue arises in the denominator of \eqref{conflux-reduction-cost}, where it implies all $p$ processors participate, which is inconsistent with the 1D decomposition depicted in \figref{conflux-1d-decomp}. Instead, at most $(p^{1/2}_1 c)$ processors are involved. This misrepresentation is evident in original code \footnote{In the code, iteration count is denoted by $k$, whereas the paper uses $t$.}, specifically for iteration $k$, where only processors with $\id{id} = [\id{pi}, \id{pj} = k \% \id{Py}, \id{pk}]$ are active (refer to \figref{conflux-panel-reduction}). The processor grid used in original code (refer to \figref{conflux-processor-grid}) indicates a configuration of $\sqrt{p} \times \sqrt{p} \times 1$ or $\sqrt{p/2} \times \sqrt{p/2} \times 2$, limiting participation to $O(\sqrt{p})$ processors. Furthermore, it is important to note that the caption of Figure 7 in the original paper (refer to \figref{conflux-cap-fig-7}) implies that the configuration for the \co{} setting should be ``$p^{1/2}_1 \times p^{1/2}_1 \times c = p^{1/3} \times p^{1/3} \times p^{1/3}$'' because $c = p^{1/3}$ when $\mM \ge n^2/p^{2/3}$. Therefore, there is a discrepancy between their code configuration and the \co{} configuration described in the paper. Nevertheless, the corrected formula that accurately reflects the involvement of processors is as follows:

\begin{align}
\frac{(n - tv)vc}{p^{1/2}_1 c} = \frac{(n - tv) v}{p^{1/2}_1} \label{eq:conflux-reduction-actual-cost}
\end{align}

The cumulative bandwidth cost throughout all iterations of  ``steps 1 and 5'' is then :

\begin{align}
\sum_{t=1}^{n/v} \frac{(n-tv) v c}{p^{1/2}_1 c} &= \sum_{t=1}^{n/v} \frac{(n-tv) v}{p^{1/2}_1} = O(n^2/p^{1/2}_1) \label{eq:conflux-reduction-actual-cumulative-cost}
\end{align}

This cost can be simplified as $O(n^2/p^{1/2})$ in accordance with their code configuration when $p^{1/2}_1 = O(p^{1/2}$), or as $O(n^2/p^{1/3})$ according to their paper configuration when $p^{1/2}_1 = p^{1/3}$. It is essential to note that both of these adjusted bandwidth costs are asymptotically greater than the cumulative costs of the remaining algorithmic steps, thereby precluding any overlap with the remaining costs. Consequently, these adjusted costs establish a lower bound for the overall bandwidth cost of the \func{COnfLUX} algorithm, specifically $\Omega(n^2/p^{1/2})$ or $\Omega(n^2/p^{1/3})$.

The remaining cost of the \func{COnfLUX} algorithm can be inferred as follows. Based on the implication of the caption of Figure 7 in original paper (refer to \figref{conflux-cap-fig-7}), we have $p^{1/2}_1 = c = p^{1/3}$ and $\mM \geq n^2/p^{2/3}$. Substituting these parameters into the bound of original Lemma 8 (refer to \figref{conflux-lemma-8}), we obtain a bandwidth cost for the remaining steps of the algorithm of $n^3/(p\sqrt{\mM}) + O(n^2/p) = O(n^2/p^{2/3})$, which is asymptotically smaller than either $O(n^2/p^{1/2})$ or $O(n^2/p^{1/3})$ as derived earlier for the $A_{10}$ and $A_{01}$ regions.

In summary, the \func{COnfLUX} algorithm incurs a communication bandwidth cost of at least $\Omega(n^2/p^{1/2})$ or $\Omega(n^2/p^{1/3})$, surpassing the claimed bound in the original paper (refer to \figref{conflux-lemma-8}). This issue primarily stems from their utilization of a 1D decomposition approach for the $A_{10}$ and $A_{01}$ regions (related to panel factorization and \proc{TRSM}), which fails to fully harness the potential of \emph{all $p$} processors for efficient communication during the reduction process. The calculation of original paper erroneously distributes the bandwidth cost across all $p$ processors, rather than the actual $(p^{1/2}_1 c)$ processors actively involved.

In addition to the above primary concern, two secondary issues are noted in the upper bound analyses:

\begin{enumerate}
\item In Lemma 8 of the original paper (\figref{conflux-lemma-8}), they assert a bandwidth cost of $n^3/(p\sqrt{\mM}) + O(n^2/p)$ for the \func{COnfLUX} algorithm. However, even following the calculation of cost in their ``steps 1 and 5'' (\figref{conflux-io-steps-1-5}), we have $\frac{2(n-tv)v\mM}{n^2} = O(\mM)$. It's worth mentioning that $O(\mM)$ can potentially exceed $O(n^2/p)$ or even the first term $n^3/(p\sqrt{\mM})$ when $n^2 < O(p^{2/3} \mM^{1/2})$.

\item Once again, in ``steps 1 and 5'' of Sect. 7.4 (\figref{conflux-io-steps-1-5}), they assert that ``\underline{with high probability}, pivots are evenly distributed among all processors'' without providing a formal proof to support this assertion. 
\end{enumerate}

\section{Methodological Concerns in the Empirical Study}
The original paper presents experimental results to validate the theoretical claims. However, upon careful examination of their code base, it becomes evident that they only tested processor grid configurations of $\sqrt{p} \times \sqrt{p} \times 1$ and $\sqrt{p/2} \times \sqrt{p/2} \times 2$ (refer to \figref{conflux-processor-grid}). Regrettably, they did not evaluate the \co{} setting of $p^{1/3} \times p^{1/3} \times p^{1/3}$ as indicated in their paper (refer to \figref{conflux-cap-fig-7}). This discrepancy has the potential to raise doubts about the empirical validity of their proposed algorithm.

\section{Issues in the Lower Bound Derivation}

The derivation of the lower bound in the original paper may also give rise to some questions. In original Lemma 7 (\figref{conflux-lemma-7}), the authors establish a parallel I/O lower bound of \eqref{conflux-lb} for a CDAG with $|V|$ vertices and $p$ processors, each equipped with $\mM$ pebbles, where $\rho$ represents the maximum computational intensity (independent of $p$), as follows.

\begin{align}
Q \geq \frac{|V|}{p \cdot \rho} \label{eq:conflux-lb}
\end{align}

While this lower bound is not necessarily invalid, the derivation itself may oversimplify the matter. A counterargument arises from the observation that in parallel computation, the total number of I/O operations typically increases proportionally to $p$, which is often asymptotically larger than in sequential computation. Unfortunately, this crucial consideration is not taken into account in the inequality \eqref{conflux-lb}. Another issue lies in the denominator of \eqref{conflux-lb}, as not all processors may be involved in every step of the computation at all times, as evidenced in the previous analysis of the upper bound of \func{COnfLUX}. Consequently, a simple division by $p$ may not yield a tight lower bound.

 \secput{concl}{Conclusion}

In this paper, we have presented a comprehensive technical reexamination of the \func{COnfLUX} algorithm, encompassing its upper bound, empirical study methods, and lower bound as documented in the work of Kwasniewski et al. \cite{KwasniewskiKaBe21}. Our analysis has brought to light several potential issues that may impact the validity of the original work's assertions. We believe that our findings contribute to a deeper comprehension of the \func{COnfLUX} algorithm and the inherent challenges involved in developing optimized matrix factorization algorithms. Our intention is to stimulate further research and foster meaningful discussion within this domain.

\section*{Acknowledgement}

We extend our sincere gratitude to the authors of the original paper for their valuable contributions to the field. We acknowledge the opportunity to engage in a constructive dialogue and offer our critique. It is important to emphasize that our objective is not to undermine the significance of their research, but rather to actively participate in the ongoing scientific discourse and collectively strive for advancements in the performance and efficiency of matrix factorization algorithms.

\bibliographystyle{IEEEtran}
\bibliography{papers}

\begin{thebibliography}{1}
\providecommand{\url}[1]{#1}
\csname url@samestyle\endcsname
\providecommand{\newblock}{\relax}
\providecommand{\bibinfo}[2]{#2}
\providecommand{\BIBentrySTDinterwordspacing}{\spaceskip=0pt\relax}
\providecommand{\BIBentryALTinterwordstretchfactor}{4}
\providecommand{\BIBentryALTinterwordspacing}{\spaceskip=\fontdimen2\font plus
\BIBentryALTinterwordstretchfactor\fontdimen3\font minus \fontdimen4\font\relax}
\providecommand{\BIBforeignlanguage}[2]{{%
\expandafter\ifx\csname l@#1\endcsname\relax
\typeout{** WARNING: IEEEtran.bst: No hyphenation pattern has been}%
\typeout{** loaded for the language `#1'. Using the pattern for}%
\typeout{** the default language instead.}%
\else
\language=\csname l@#1\endcsname
\fi
#2}}
\providecommand{\BIBdecl}{\relax}
\BIBdecl

\bibitem{KwasniewskiKaBe21}
G.~Kwasniewski, M.~Kabic, T.~Ben-Nun, A.~N. Ziogas, J.~E. Saethre, A.~Gaillard, T.~Schneider, M.~Besta, A.~Kozhevnikov, J.~VandeVondele, and T.~Hoefler, ``On the parallel i/o optimality of linear algebra kernels: near-optimal matrix factorizations,'' in \emph{SC '21: Proceedings of the International Conference for High Performance Computing, Networking, Storage and Analysis}, 2021, p.~70.

\bibitem{SolomonikDe11}
E.~Solomonik and J.~Demmel, ``Communication-optimal parallel 2.5d matrix multiplication and lu factorization algorithms,'' in \emph{Proceedings of the 17th International Conference on Parallel Processing - Volume Part II}, ser. Euro-Par'11.\hskip 1em plus 0.5em minus 0.4em\relax Berlin, Heidelberg: Springer-Verlag, 2011, pp. 90--109.

\end{thebibliography}

	
\end{document}